\documentclass[conference,a4paper]{APSIPA2018}
\usepackage{multirow}
\usepackage{amsmath}
\usepackage[psamsfonts]{amssymb}
\usepackage{amsxtra}
\usepackage{threeparttable}
\usepackage{booktabs,subcaption,amsfonts,dcolumn}
\usepackage{bm}
\usepackage{blindtext}
\usepackage{tikz}
\usetikzlibrary{calc} 
\usepackage{cite}
\usepackage{adjustbox}
\usepackage{tabularx}
\usepackage{algcompatible}

\usepackage{algorithm}
\usepackage{algpseudocode}

\usepackage{stfloats}
\usepackage{lipsum}

\begin{document}

\title{Unsupervised Pattern Discovery from Thematic Speech Archives Based on Multilingual Bottleneck Features}

\author{%

Man-Ling Sung, Siyuan Feng and
Tan Lee \\

Department of Electronic Engineering, \\
The Chinese University of Hong Kong, 
Shatin, Hong Kong SAR, China\\

E-mail: \{mlsung, tanlee\}@ee.cuhk.edu.hk, siyuanfeng@link.cuhk.edu.hk

}

\maketitle
\thispagestyle{empty}

\begin{abstract}
The present study tackles the problem of automatically discovering spoken keywords from untranscribed audio archives without requiring word-by-word speech transcription by automatic speech recognition (ASR) technology. The problem is of practical significance in many applications of speech analytics, including those concerning low-resource languages, and large amount of multilingual and multi-genre data. We propose a two-stage approach, which comprises unsupervised acoustic modeling and decoding, followed by pattern mining in acoustic unit sequences. The whole process starts by deriving and modeling a set of subword-level speech units with untranscribed data. With the unsupervisedly trained acoustic models, a given audio archive is represented by a pseudo transcription, from which spoken keywords can be discovered by string mining algorithms. For unsupervised acoustic modeling, a deep neural network trained by multilingual speech corpora is used to generate speech segmentation and compute bottleneck features for segment clustering. Experimental results show that the proposed system is able to effectively extract topic-related words and phrases from the lecture recordings on MIT OpenCourseWare.

\end{abstract}


\section{Introduction}

In recent years, automatic speech recognition (ASR) technology is advancing to the level that is considered adequate for daily use in human-computer interaction \cite{Saon2017,Hori2017}. The high performance level is contributed largely by the effectiveness of deep learning algorithms with large amount of training data \cite{dahl2012context,Chen2017investigating}. For mainstream commercial systems, there is no much concern on the availability of training data and linguistic knowledge about the intended languages. With fine-grained or coarse transcription for part or all of the training data, supervised approaches could be applied to the training of a deep neural network for acoustic-phonetic mapping \cite{dahl2012context,sak2014long,bahdanau2016end}. This is considered a rather straightforward process.

There are many application scenarios where transcribed training data are difficult or even impossible to acquire. One of the scenarios concerns those unpopular and less-studied languages or dialects, for examples, ethnic minority languages in China. In the ASR research community, low-resource languages (or zero-resource in the extreme case) refer to the cases that most of the key elements of linguistic knowledge required for ASR system development, e.g., definition of phonemes, pronunciation lexicon, orthographically represented data, etc., do not exist, although a certain amount of audio-form speech data may be available \cite{versteegh2015zero,dunbar2017zero}. There are also situations that the linguistic resources in existence do not adequately represent the spoken language in actual usage, for examples, regional variants of a major language, code-mixing speech, and technical language in a highly specialised area.

Another application scenario with practical significance is the unsupervised spoken term discovery from large-scale multi-genre audio archives. By multi-genre, we refer to a high diversity of audio content, including speech of different speaking styles and accents, and from varying acoustic channels, and all kinds of non-speech sounds. Examples are publicly available broadcast media content, live recordings of lectures/seminars/meetings, and personal digital recordings. While improving ASR performance for multi-genre speech transcription has attracted great research interest \cite{ali2017speech}, unsupervised pattern discovery is considered a pragmatic approach to efficient indexing, search and organizing of complex audio content.

The present study tackles the problem of unsupervised keyword discovery from raw audio of topic-specific classroom lectures. The experimental dataset is adapted from unedited video recordings in the MIT OpenCourseWare, covering various courses on Mathematics, Engineering and Computer Programming \cite{MITopen}. In addition to the teacher's speech, the audio content contain students' speech (e.g., asking or responding questions), cough, laughter, chalk-writing sounds, furniture sounds, video sound, etc. The teachers may or may not be native speakers of English. Some of them carry very strong foreign accents. Our objective is to automatically find out a set of keywords that can represent the subject of a lecture or course, without performing explicit speech transcription.

To tackle this problem, we develop a two-stage approach, which comprises unsupervised acoustic modeling and decoding, followed by pattern mining in acoustic unit sequences. Subword-level acoustic models are trained from untranscribed audio recordings using an unsupervised approach. For the segmentation of audio frames and clustering of subword units, we propose to incorporate language-independent bottleneck features (BNFs) from a deep neural network (DNN) trained by multilingual speech corpora  \cite{vesely2012language}.
The multilingual DNN is expected to provide a wide coverage and rich representation of acoustic and phonetic variations, so as to better characterize the unseen speech data. With unsupervisedly trained acoustic models, each lecture recording can be decoded into a pseudo transcription, which is in the form of subword unit sequence. A spoken keyword in the lecture is identified as a distinctive subsequence pattern that occurs multiple times in the pseudo transcription. The performance of the proposed system is evaluated by comparing and relating the discovered ``keywords'' with the ground-truth transcription provided at the MIT OpenCourseWare website.






\tikzset{
        block/.style = {draw, fill=white, rectangle, minimum height=3em, minimum width=3em},
        tmp/.style  = {coordinate}, 
        sum/.style= {draw, fill=white, circle, node distance=1cm},
        input/.style = {coordinate},
        output/.style= {coordinate},
        pinstyle/.style = {pin edge={to-,thin,black}
        }
    }
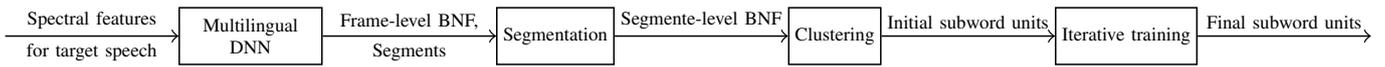
\begin{figure*}[htb]

    \begin{center}
    {\tiny
\begin{adjustbox}{width=\textwidth}
    \begin{tikzpicture}
        \matrix[column sep=1.6cm, row sep=.7cm] {%
           \node [input] (input) {}; 
           &\node [block,align=center,text width=1.2cm] (nnet) {Multilingual DNN}; 
           &\node [block] (seg) {Segmentation};
           &\node [block] (clu) {Clustering};
           &\node [block] (iter) {Iterative training};
           &\node [output] (output) {}; 
           \\
        };  
        \draw [->] (input) -- node[above] {Spectral features} node[below] {for target speech} (nnet);
        \draw [->]  (nnet) -- node[above] {Frame-level BNF,} node[below] {Segments}(seg);
        \draw [->]  (seg) -- node[above]{Segmente-level BNF}  (clu);
        \draw [->] (clu) -- node[above] {Initial subword units} (iter);
        \draw [->] (iter) -- node[above] {Final subword units} (output);
    \end{tikzpicture}
    \end{adjustbox}
    
    }
 \end{center}
\caption{Unsupervised acoustic  modeling framework used to train acoustic subword models and generate subword unit sequences for target speech.}
\label{fig:system}
\end{figure*}

\tikzset{
        block/.style = {draw, fill=white, rectangle, minimum height=3em, minimum width=3em},
        tmp/.style  = {coordinate}, 
        sum/.style= {draw, fill=white, circle, node distance=1cm},
        input/.style = {coordinate},
        output/.style= {coordinate},
        pinstyle/.style = {pin edge={to-,thin,black}
        }
    }
    
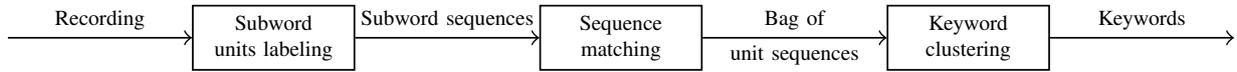
\begin{figure*}[htb]

    \begin{center}
    {\tiny
\begin{adjustbox}{width=0.9 \textwidth}
    \begin{tikzpicture}
        \matrix[column sep=1.5cm, row sep=.7cm] {%
           \node [input] (input) {}; 
           &\node [block,align=center,text width=1.2cm] (nnet) {Subword units labeling}; 
           &\node [block,align=center,text width=1.2cm] (seg) {Sequence matching};
           &\node [block,align=center,text width=1.2cm] (clu) {Keyword clustering};
           &\node [output] (output) {}; 
           \\
        };  
        \draw [->] (input) -- node[above] {Recording} (nnet);
        \draw [->]  (nnet) -- node[above] {Subword sequences} (seg);
        \draw [->]  (seg) -- node[above] {Bag of  } node[below] {unit sequences}(clu); 
        \draw [->] (clu) -- node[above] {Keywords} (output);
    \end{tikzpicture}
    
    \end{adjustbox}
    }
 \end{center}
\caption{Unsupervised pattern mining framework used to discover keywords in subword unit sequences of target speech.}
\label{fig:system2}
\end{figure*}

\section{Previous work on unsupervised acoustic modeling and pattern discovery }


\subsection{Unsupervised acoustic modeling}
\label{subsec:usm}
Unsupervised acoustic modeling
aims at finding basic speech units, which are desirably equivalent to phonemes,
from untranscribed speech data. 
Previously investigated approaches can be divided into two categories, namely \emph{bottom-up modeling} and \emph{top-down modeling}.
In the bottom-up approach, speech is viewed as a sequence of low-level components, e.g., frames or segments, which can be grouped by clustering techniques to define higher-level structures \cite{zhang2010towards,wang2015acoustic,chen2015parallel}. The learned clusters are regarded as the basic units to represent the language concerned.
Top-down modeling methods seek use of higher-level knowledge to provide constraints and guide the modeling of low-level speech components \cite{jansen2011towards,jansen2013weak}. The higher-level knowledge, e.g., word/phrase segment pairs,
could be obtained by a process known as spoken term discovery (STD). 
In recent studies, DNNs has been shown effective in improving performance of both bottom-up and top-down modeling methods
\cite{badino2014auto,kamper2015unsupervised,renshaw2015comparison,chen2017multilingual}. 

The commonly used bottom-up framework for unsupervised acoustic modeling consists of three steps, namely,
speech segmentation, segment clustering, and iterative modeling. Speech segmentation aims to divide a speech utterance into variable-length segments based on, for instance, the spectral discontinuities. 
Segmentation could be data-driven, e.g. the bottom-up hierarchical clustering method presented in \cite{QiaoShimomuraMinematsu2008}, or based on out-of-domain knowledge and/or resources, e.g., language-mismatched phone recognizers \cite{feng2016exploiting}.
Segment clustering is to find and group speech segments that share similar acoustic properties. 
Various clustering algorithms such as Gaussian mixture model (GMM) \cite{WangLeungLeeEtAl2012}, segmental GMM (SGMM) \cite{GishNg1993}, vector quantization (VQ) \cite{LeeSoongJuang} and spectral clustering \cite{wang2015acoustic,feng2016exploiting}, have been investigated. 
With the cluster labels as initial tokenization, acoustic models are trained and refined by iteratively performing model parameter estimation and decoding, in a supervised manner. Following the terminology of \cite{LeeSoongJuang}, this framework is referred to as acoustic segment modeling (ASM) in this study.

\subsection{Unsupervised pattern discovery}

Unsupervised pattern discovery (also known as spoken term discovery (STD) \cite{dunbar2017zero}, or unsupervised term discovery (UTD) \cite{lyzinski2015evaluation}) refers to the task of automatically
discovering words and linguistic entities from audio archives without supervision \cite{park2008unsupervised}.
Unsupervised pattern discovery from audio signals could be done in two steps \cite{park2008unsupervised,zhang2009unsupervised}: (1)
identifying similar audio segments from target audio archives, and
(2) clustering the segments into groups of discovered patterns.
In \cite{park2008unsupervised}, Park and Glass proposed a segmental dynamic time wrapping (DTW) algorithm to discover similar audio patterns from pairs of utterances by comparing at acoustic level. They further applied graph-based clustering towards similar segments, in order to find the
most common word-/phrase-like patterns. A number of extensions to \cite{park2008unsupervised} were made mainly to improve the feature representations, 
e.g., using Gaussian posteriorgram \cite{zhang2009unsupervised}, language-mismatched phoneme posteriorgram and unsupervised subword posteriorgram \cite{wang2012acoustic}. Another direction of improvement was to improve the computation efficiency \cite{jansen2011efficient}. 

Recently, there are works focusing on the topic of \emph{lexicon discovery} from untranscribed speech \cite{lee2015unsupervised,kamper2017segmental}. It is closely related to unsupervised pattern discovery, but emphasizing on full-coverage segmentation of speech into word-like units.

\section{Unsupervised acoustic modeling and decoding}
\label{sec:uam}
In the present study, the ASM framework is adopted for unsupervised acoustic modeling and decoding.
The novelty of our work lies in that a multilingual DNN trained with resource-rich language resources is involved in both speech segmentation and segment representation. The proposed system design is shown as in Fig. \ref{fig:system}.


\subsection{Segmentation and BNF extraction}
\label{ssec:SegBNF}
\begin{figure}[t]
\begin{center}
\begin{tikzpicture}[scale=0.7]
\node[text width=3cm] at (-0.15,2) {input};
\filldraw[fill=gray,draw=black] (-1,0.5) rectangle (-0.6,3.5);
\draw (-0.6,3.5) -- (0,4) -- (0,0) -- (-0.6,0.5);
\filldraw[fill=gray,draw=black] (0,0) rectangle (0.4,4);
\draw (0.4,4) -- (1,0) -- (1,4) -- (0.4,0);
\draw (1.4,4) -- (2,0) -- (2,4) -- (1.4,0);
\filldraw[fill=gray,draw=black] (1,0) rectangle (1.4,4);
\filldraw[fill=gray,draw=black] (2,0) rectangle (2.4,4);
\draw (2.4,4) -- (3,2.8) -- (3,1.2) -- (2.4,0);
\node[text width=3cm] at (3.4,4.9) {bottleneck layer};
\draw [->,dotted,thick,>=stealth] (3.2,4.5) -- (3.2,2.8);

\filldraw[fill=gray,draw=black] (3,1.2) rectangle (3.4,2.8);
\draw (3.4,1.2) -- (4,0) -- (4,4) -- (3.4,2.8);
\filldraw[fill=gray,draw=black] (4,0)  rectangle (4.4,4);

\filldraw[fill=gray,draw=black] (6,4.1) rectangle (6.4,5.5);
\filldraw[fill=gray,draw=black] (6,2.1) rectangle (6.4,3.5);
\filldraw[fill=gray,draw=black] (6,-0.9) rectangle (6.4,0.5);
\node[text width=3cm] at (8.2,3.8) {language 1};
\node[text width=3cm] at (8.2,1.8) {language 2};
\node[text width=3cm] at (8.2,-1.2) {language N};

\draw[loosely dotted, very thick] (6.2,1.6) -- (6.2,0.8);

\draw (4.4,0) -- (6,4.1) -- (6,5.5) -- (4.4,4);
\draw (4.4,0) -- (6,2.1) -- (6,3.5) -- (4.4,4);
\draw (4.4,0) -- (6,-0.9) -- (6,0.5) -- (4.4,4);
\end{tikzpicture}
\end{center}

\caption{Multilingual bottleneck network used to obtain language independent acoustic information.}
\label{fig:nnet}
\end{figure}
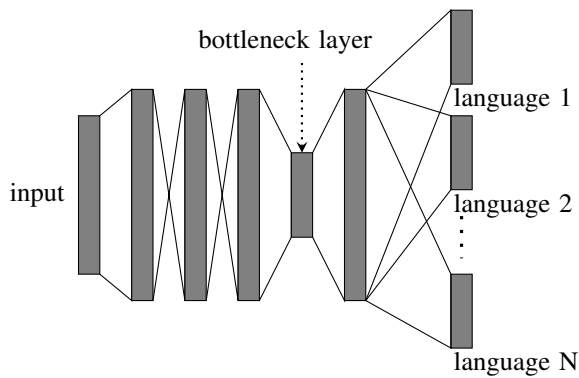

The input utterance is divided into variable-length segments, from which fixed-dimension feature representations are extracted. In the proposed system, a multilingual DNN is trained for this purpose. The architecture of a multilingual DNN is illustrated as in Figure \ref{fig:nnet}.
It contains a set of language-specific output layers and a number of shared hidden layers, including a low-dimensional bottleneck (BN) layer. Training of the DNN follows stochastic gradient descent (SGD) algorithm, to minimize language-weighted average cross-entropy loss function \cite{huang2013cross}.  

Given an input utterance, the trained DNN would produce multiple sets of phoneme-level time alignments at the block-softmax layers. These hypothesized alignments are combined to give a single segmentation as described in Feng et al. \cite{feng2016exploiting}, i.e., hypothesized segment boundaries within an internal of $20$ ms are merged. 
Frame-level BNF of an input utterance are extracted from the bottleneck layer. Segment-level BNFs are obtained by averaging the frame-level features.

\subsection{Segment clustering}
\label{ssec:cluster}
\begin{figure}[t]
\begin{center}
\includegraphics[scale=0.25]{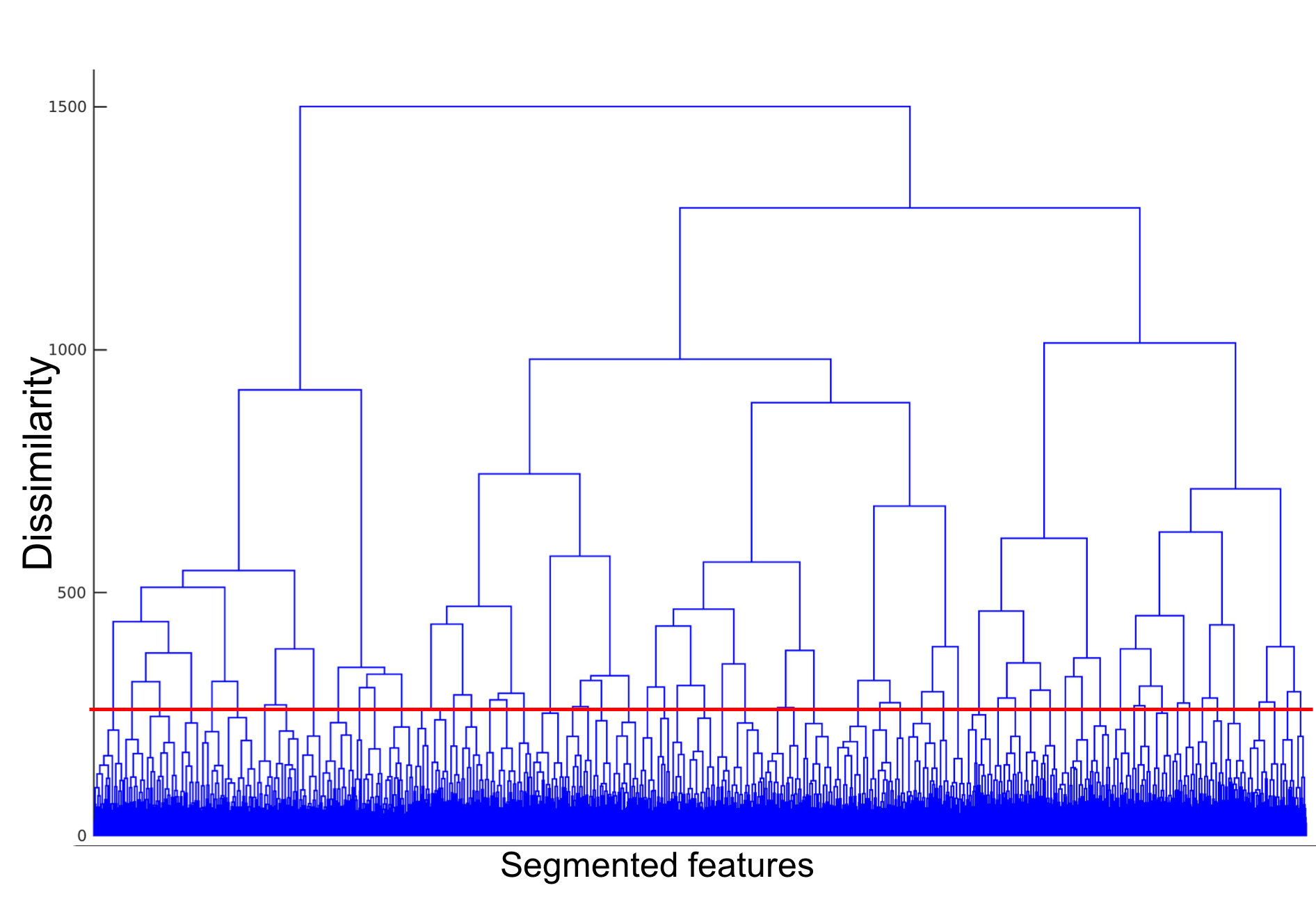}
\end{center}
\caption{Dendrogram showing hierarchical clustering of 100,000 segments.}
\label{fig:dendro}
\end{figure}

The BNFs obtained from all segments have the same dimension. By a BNF clustering process, segments with similar acoustic properties are grouped together. A practical issue for segment clustering is that the number of clusters is unknown, as in most cases linguistic knowledge on the concerned language is completely unavailable. 
The hierarchical agglomerative clustering (HAC) \cite{ward1963hierarchical} approach is applied to obtain a general 
picture on the similarity among segments, and allow the number of clusters to be determined manually. In this study,
the Ward's minimum variance \cite{murtagh2011ward} is used as the linkage criterion for HAC. An example dendrogram describing the result of HAC on a lecture recording from MITOpenCourseWare (refer to Section \ref{sec:experiment} for details) is illustrated as in Fig. \ref{fig:dendro}.

After determining the number of clusters $R$, the segments from the entire dataset are clustered by the $k$-means algorithm. Each of the resulted cluster is assigned a specific label that is used to represent all its members. The clusters are expected to correspond to a set of linguistically relevant speech units are subword level. Given an input utterance, the sequence of segment labels is regarded as a kind of \emph{pseudo transcriptions}, which can be used for further analysis and comparison.

\subsection{Iterative training and decoding}
\label{ssec:training}
The availability of pseudo transcription makes supervised training of acoustic models possible. Specifically, DNN-HMM acoustic models \cite{dahl2012context} can be trained to represent the unsupervisedly learned subword units, and these models are refined in conjunction with the pseudo transcriptions of all training data in an iterative manner, as elaborate below:
\begin{enumerate}
\item Train an initial set of DNN-HMM acoustic models with the pseudo transcriptions resulted from segment clustering.
\item Decode the training data with the current models and obtain new pseudo transcriptions\label{enum:step2}.
\item Train the acoustic models with the new pseudo transcriptions\label{enum:step3}.
\item Repeat Step \ref{enum:step2}) and \ref{enum:step3}) until convergence.
\end{enumerate}
Iterative training is carried out with the domain-specific dataset, i.e., all lectures from the same course. In this way, acoustic models and pseudo transcriptions are jointly optimized for the target application.

After iterative training, the final version of pseudo transcriptions, in the form of subword unit sequences, are used for the subsequent process of keyword discovery.

\section{Pattern Discovery from Subword Unit Sequence}
\label{ssec:pattern}

With the unsupervisedly trained acoustic models, the audio content of a lecture can be represented by a pseudo transcription (subword unit sequence). A keyword spoken frequently during the lecture is identified by a distinctive pattern that repeatedly occurs in the transcription. The process of searching for such candidate patterns consists of two parts: 
\begin{itemize}
\item Sequence matching between each pair of pseudo transcriptions and generating ``bags of unit sequences'';
\item Clustering the ``bags of unit sequences'' into groups of similar sequence patterns, which correspond to keywords. 
\end{itemize}

\subsection{Generating ``bags of unit sequences''}
\label{ssec:bag_of_seq}
Identification of similar or closely related subsequences in long symbol sequences is an important problem in bioinformatics. In \cite{smith1981identification}, an algorithm of inexact matching between a pair of short symbol sequences was described. Based on this algorithm, the ``bags of unit sequences''  containing only the matching subsequences are generated. This algorithm is referred to as \emph{local sequence alignment} in this paper.
Pseudo code for this algorithm is listed in Algorithm \ref{LocalAlign}. 

In our application, symbol sequences are pseudo transcriptions, i.e. subword unit sequences, acquired from unsupervised acoustic modeling, as discussed in Section \ref{sec:uam}. 
The motivation for allowing inexact matching of ``bags of unit sequences'' is to cope with the possible decoding errors in pseudo transcriptions and pronunciation variations. 
Suppose there are $N$ utterances for the target untranscribed speech. Each time we pick up $2$ symbol sequences $A$ and $B$, corresponding to $2$ utterances, and perform Algorithm \ref{LocalAlign} towards $A$ and $B$ to generate ``bags of unit sequences''. After $\frac{N(N-1)}{2}$ times, all possible pairs of sequences are used to perform Algorithm \ref{LocalAlign}. Finally, all the discovered ``bags of unit sequences'' are stored, and will be used for subsequent clustering process.  
We are only interested  in the ``bags of unit sequences'' with $4$ symbols or longer, to make sure they mainly cover content-related words.


\begin{algorithm}
  \caption{Local sequence alignment}\label{LocalAlign}
  \begin{algorithmic}[1]
    \Procedure{LocalAlign}{$A = a_1 a_2 ... a_n, B = b_1 b_2 ... b_m$}
      \State $s(a_i,b_j)= \begin{cases}
    +1&\text{, if $a_i,b_j$ match}\\
    -1&\text{, if $a_i,b_j$ mismatch}
    \end{cases}$  \Comment{Similarity score between sequence elements $a_i$ and $b_j$}
    
        \State Compute an $(n+1) \times (m+1)$ matrix $\mathbf{P}$, where the element $p_{i, j}$ is, 
        \Statex $p_{i, j} = \begin{cases}
		0&\text{, $i$ or $j = 0$}\\
		\max 
		\begin{cases}
		p_{i-1, j-1} + s(a_i, b_j)\\
		p_{i-1, j}\\
		p_{i, j-1}\\
		0
		\end{cases}&\text{, elsewhere}
		\end{cases}$  
      \State Traceback from $p_{n,j^*}$ ending with an element of $\mathbf{P}$ equal to $0$, where $p_{n,j^*}$ are local maxima of $\{p_{n,j} | 0 \leq j \leq m \}$, to obtain common subsequence pairs in $A$ and $B$. 

      \State Store all obtained subsequences with reasonable length into the ``bag of words''.

     \EndProcedure
  \end{algorithmic}
\end{algorithm}


\subsection{Sequence clustering}
\label{subsec:seq_clus}
The ``bag of unit sequences" created as in Section \ref{ssec:bag_of_seq} contain a large number of unit sequences of different lengths. These sequences are clustered into groups using the leader clustering algorithm  \cite{hartigan1975clustering}, as depicted in  Algorithm \ref{leader}.
The radius of each cluster is given as $T$. To avoid clusters from overlapping to much, we set the minimum distance between centroids into $a*T$, where $a$ is larger than $0$.
Leader clustering is sensitive to the initialization of centroids.
To avoid having the centroid that is not the representative of the cluster due to poor initialization (e.g. outliers), the centroid is updated with the most representative sequence, i.e., the sequence having the least total distance with all cluster members, as is shown in Line \ref{algo:update_centroid} of Algorithm \ref{leader}. 
The clustering process iterates until the number of clusters converges (remain unchanged).


The normalized Levenshtein distance $||L(x,y)||$ used in Algorithm \ref{leader} is defined as,
\begin{equation}
\label{eqt:lev_dis}
||L(x,y)|| = b \dfrac{L(x,y)} {\sqrt{|x|^2 + |y|^2}},
\end{equation}
where $L(x,y)$ is the Levenshtein distance \cite{Vreda1999Lev} between sequences $x$ and $y$. $b$ is an adjustable scalar.
The normalization allows clusters to be formed with consideration of sequence length: short sequences to have more strict match (e.g. $1$ difference out of $5$), and longer sequence to have looser match ($2$-$3$ differences out of $10$).


\begin{algorithm}
  \caption{Leader clustering}\label{leader}
  \begin{algorithmic}[1]
    \Procedure{Leader}{bag of sequences $S$}
      \State Initial a point $i$ to $\mathrm{centroid}$
      \For{each point $p$ in $S$ }
        \If{$||L(i,p)|| > a*T$ for all $i$ in $\mathrm{centroid}$, $a>0$:}
        \State Add $p$ to $\mathrm{centroid}$
        \EndIf
      \EndFor
        \State 	Assign each point $p$ in $S$ to its closest cluster $i$ with $||L(i,p)|| < T $.
        \For{each group}
        \State Update $\mathrm{centroid}$ with the representative of the cluster (measured by smallest total distance with same group members).\label{algo:update_centroid}
        \EndFor
      \State Repeat steps 2-11 until the number of clusters converge.
    \EndProcedure
  \end{algorithmic}
\end{algorithm}

\section{Experimental setup}
\label{sec:experiment}
 
\subsection{Training data}
\label{ssec:data}

The proposed system is evaluated on the task of unsupervised keyword discovery from audio recordings of academic lectures. The audio data are extracted from three courses that are publicly available at the MIT OpenCourseWare website \cite{MITopen}. The courses are ``Mathematics for Computer Science'' (MATH), ``Principles of Digital Communication II'' (COMM) and ``Introduction to Computer Science and Programming in Python'' (PYTH). The recording conditions of the three courses were similar. Each course consists of $12 - 25$ lectures. The duration of lecture is in the range of $45-70$ minutes. The course teacher of MATH spoke with French accent. The speaking rate in PYTH was relatively fast and that in COMM was slow. The long recordings are divided into short segments (roughly $5 - 10$  seconds) for further processing. 



\subsection{System set-up}
\label{ssec:system}

Unsupervised acoustic modeling is carried out for the three courses separately. That is, a set of subword units are learned from all lecture recordings of each course, leading to a set of acoustic models tailored for the course. 

A multilingual DNN with BN layer is trained with $5$ speech corpora of $4$ resource-rich languages: TIMIT (English) \cite{timit}, WSJ (English) \cite{paul1992design}, CUSENT (Cantonese) \cite{lee2002spoken}, 863 (Mandarin) \cite{qian2004introduction} and a distant-speech database of German \cite{radeck2015open}, using Kaldi \cite{povey2011kaldi}. The DNN has $5$ hidden layers with dimensions of $1500$, $1500$, $1500$, $40$ (BN layer) and $1500$ respectively. There are $5$ block-softmax output layers, corresponding to the $5$ speech corpora. The state-level labels required for supervised training of the multilingual DNN are obtained from the context-dependent GMM-HMM (CD-GMM-HMM) trained separately for the $5$ corpora. $23$-dimensional Mel-scale filter-bank features (fbanks) are extracted for
training both CD-GMM-HMM and DNN.

The trained multilingual DNN is used to obtain subword-level segmentation and generate frame-level BNFs for the lecture recordings (see Section \ref{ssec:SegBNF}). The segmentation is derived from the $5$ block-softmax layers. Segment-level BNFs are obtained by averaging the frame-level features. By applying $k$-means clustering to segment-level features, $55$ clusters are obtained, each denoting a subword-level unit. The number of clusters is determined by HAC, as discussed in Section \ref{ssec:cluster}.

For iterative training, acoustic models are trained in a course-specific manner, also using Kaldi \cite{povey2011kaldi}\footnote{\fontfamily{pcr}\selectfont kaldi/egs/timit/s5/run.sh}. In each iteration, GMM-HMM acoustic models with speaker adaptive training (GMM-HMM-SAT) are trained beforehand to generate feature-space maximum likelihood linear regression (fMLLR) features, with the latest version of pseudo transcriptions and multilingual BNFs as input features. The fMLLRs and HMM state-level alignments generated by GMM-HMM-SAT are used to train the DNN-HMM acoustic models, followed by decoding target speech to obtain an updated version of pseudo transcriptions, i.e., the $1$-best path of decoding lattices. The language model needed for decoding is trained with exact the same version of transcriptions used for GMM-HMM-SAT training in this iteration.
During iterative training of course-specific acoustic models and decoding target data into pseudo transcriptions,
the frame-level unit labels before and after each iteration are compared. If the percentage difference is below $0.1\%$, convergence of training is assumed. In our experiments, it is observed that $5$ to $6$ iterations are needed for convergence.

As a comparison, iterative training with the same configuration except for replacing multilingual BNFs with fbanks (as inputs to GMM-HMM-SAT training) is also tested.
The system with multilingual BNFs converges faster than that with the fbanks\footnote{Due to limited space, experimental results on comparing BNFs and fbanks are not presented here.}.


After iterative training, the recordings are represented by pseudo transcriptions. Bag of subword sequences, generated by performing Algorithm \ref{LocalAlign}, are clustered by Algorithm \ref{leader} in order to obtain similar sequence patterns, i.e. the discovered keywords. The parameter $b$ in Equation (\ref{eqt:lev_dis}) is set to $4$. We experimented various parameters from $T: 0.7 - 1.6$ (with interval $0.1$), $a: 1 - 2$ (with interval $0.1$), and found out that $T = 1.4$, $a = 1.8$ could lead to relatively good performance in terms of cluster number, purity and  cluster size.



\section{Results and Analysis}

Given a set of lecture recordings, the proposed system is able to generate a certain number of clusters of subword unit sequences. Unit sequences in the same cluster should be similar (based on the normalized Levenshtein distance) and are expected to represent the same word or phrase being spoken in the lectures. For the intended task of keywords discovery, it is desired that a significant portion of the content-related words could be covered by the unsupervisedly generated clusters. In this section, we analyze the automatically discovered keyword clusters with respect to the frequently occurred words and phrases in the ground-truth transcriptions (available at the MIT OpenCourseWare website\footnote{\fontfamily{pcr}\selectfont ocw.mit.edu/index.htm}). 

\subsection{Quality of discovered ``word'' clusters}
\label{ssec: cluster_analysis}

We examine the clustering results for $2$ selected lectures in the course MATH. For Lecture 4 (``Number Theory I'', $80$ min. long), the system generates $95$ keyword clusters from $34,313$ candidate unit sequences. For Lecture 8 (``Graph Theory II: Minimum Spanning Trees'', $83$ min. long), there are  $119$ clusters from $25,899$.

Tables \ref{lec_4} and \ref{lec_8} list the words corresponding to the $10$ longest sequence clusters in the two lectures, respectively. It is observed that most of these clusters correspond to words that are related to the lecture topic. Clusters of sequences with $12$ to $16$ subword units generally have high purity. A cluster with high purity provides a valid representation of a specific word or phrase. As the sequence length decreases, the cluster's purity tends to decrease. Sequences with less than $5$ units typically correspond to parts of different words that have similar pronunciations, e.g., ``so'', ``(al)so'', and ``so(lve)''; ``in'' and ``in(teger)''.

\newcolumntype{L}{>{\arraybackslash}m{4.5cm}}
\newcolumntype{C}{>{\arraybackslash}m{0.8cm}}
\begin{table}[t]
\centering

\caption{Lecture 4: Number Theory I}
\label{lec_4}
\small
\begin{tabular}{|l|L|C|C|}
\hline
Cluster \# & Corresponding words & Cluster size & Purity                     \\\hline
58         & divide any result                       & 2   & 100\%                       \\\hline
63         & the zero steps                        & 2   & 100\%                       \\\hline
70         & Multiple of                          & 2  & 100\%                       \\\hline
56         & linear combination                      & 17   & 100\%                       \\\hline
40         & the/a number theory                             & 5    & 100\%                       \\\hline
70         & a and b & 14       & 100\% \\\hline
28         & use the lemma again                                           & 2     & 100\%                       \\\hline
43         & Greatest common, The greatest common                     & 18     & 100\%                       \\\hline
77 & *chalk-writing sounds*  & 8   & 100\%                       \\\hline
76  & gallon jug & 15   & 100\%                       \\\hline
\end{tabular}
\end{table}


\newcolumntype{L}{>{\arraybackslash}m{4.5cm}}
\newcolumntype{C}{>{\arraybackslash}m{0.8cm}}
\begin{table}[t]
\centering
\caption{Lecture 8: Graph Theory II: Minimum Spanning Trees}
\label{lec_8}
\small
\begin{tabular}{|l|L|C|C|}

\hline
Cluster \# & Corresponding words                               & Cluster size & Purity     \\\hline
70         & b equal to                              & 2     & 100\%      \\\hline
66         & this particular edge, this particular err             & 4   & 75\%       \\\hline
57         & connected subgraph, the subgraph         & 7     & 100\%      \\\hline
89         & A subgraph,The smaller part                       & 4      & 50\%, 50\% \\\hline
87         & still connected, both connected                     & 4       & 100\%      \\\hline
83         & Double star is                                    &  2       & 100\%      \\\hline
80  & So we know that  & 2   & 100\%      \\\hline
116  & Vertices, vertices that, -ices have       &  7     & 85.7\%    \\\hline
45   & the spanning tree, a spanning tree, spanning tree & 25  & 100\%      \\\hline
109        & vertices  &  4  & 100\%      \\\hline 
\end{tabular}
\end{table}

The same word may be represented by more than one clusters. For example, clusters \#116 and \#109 for Lecture 8 (Table \ref{lec_8}) both correspond to ``vertices''. It is also observed that some of discovered words can be composed by other clusters of shorter sequences. Some of the clusters represent non-speech sounds, e.g., cluster \#77 in Table \ref{lec_4} for ``chalk-writing sounds'', which are very common in live recordings of lectures.


\subsection{Coverage of discovered words}

In this section, the coverage of automatically discovered ``words'' is examined with respect to the ground-truth transcriptions. For the three courses that we are experimenting with, word-by-word speech transcriptions are available at the the course website. Word-level trigrams, bigrams and unigrams are computed from the transcription for each lecture session or all lectures in a course, with the function words like ``is'', ``a'', ``the'' being discarded.

\newcolumntype{C}[1]{>{\centering\let\newline\\\arraybackslash\hspace{0pt}}m{#1}}

\begin{table}

\caption{Matching results for the lecture of ``Object-Oriented Programming''}
\begin{subtable}[t]{0.5\textwidth}\centering
\begin{tabular}{|C{3.5cm}|C{1.5cm}|C{1.5cm}|}
\hline
\textbf{Trigram}  & \textbf{Count} & \textbf{Match ?}\\
\hline
a coordinate object   & 14    & yes    \\
can interact with     & 13    & yes    \\
a fraction object     & 9     & partly \\
an object of          & 7     & yes    \\
of the class          & 7     & yes    \\
of type coordinate    & 7     & partly \\
the exact same        & 6     &        \\
you can create        & 6     &        \\
is equal to           & 6     &        \\
going to define       & 6     & yes    \\\hline
\end{tabular}
\caption{\footnotesize Trigram}
\label{tab:table1_a}
\end{subtable}

\begin{subtable}[t]{0.5\textwidth}\centering
\begin{tabular}{|C{3.5cm}|C{1.5cm}|C{1.5cm}|}
\hline
\textbf{Bigram}  & \textbf{Count} & \textbf{Match ?}\\
\hline
an object             & 25    & yes    \\
the class             & 22    & yes    \\
coordinate object     & 20    & partly \\
interact with         & 19    & yes    \\
a list                & 17    &        \\
a coordinate          & 17    & yes    \\
 the object            & 14    & yes    \\
fraction object       & 14    & partly \\
object that           & 14    & yes    \\
the x                 & 13    &        \\
this method           & 13    & yes    \\
can interact          & 13    & yes    \\
create a              & 13    &        \\
  a fraction            & 12    & yes    \\
data attributes       & 12    & yes    \\
of type               & 11    & yes    \\
for example           & 11    & yes    \\
to define             & 10    &        \\
the list              & 10    &        \\
underscore underscore & 10    &        \\
\hline
\end{tabular}
\caption{\footnotesize Bigrams}
\label{tab:table1_a}
\end{subtable}

\begin{subtable}[t]{0.5\textwidth}\centering
\begin{tabular}{|C{3.5cm}|C{1.5cm}|C{1.5cm}|}
\hline
\textbf{Unigram} & \textbf{Count} & \textbf{Match ?} \\\hline
object                & 118   & yes    \\
coordinate            & 70    & yes    \\
class                 & 60    & yes    \\
type                  & 47    & yes    \\
method                & 46    & yes    \\
data                  & 45    & yes    \\
objects               & 44    & yes    \\
right                 & 38    &        \\
i                     & 35    &        \\
list                  & 32    &        \\
python                & 30    & yes    \\
x                     & 29    &        \\
create                & 28    &        \\
self                  & 28    &        \\
print                 & 27    &        \\
one                   & 25    & yes    \\
c                     & 24    & yes    \\
dot                   & 24    &        \\
attributes            & 22    & yes    \\
fraction              & 21    & yes    \\
underscore            & 20    &        \\
value                 & 20    & yes    \\
interact              & 19    & yes    \\
y                     & 18    &        \\
define                & 18    &        \\
init                  & 17    &        \\
add                   & 17    &        \\
lists                 & 16    &        \\
car                   & 16    &        \\
code                  & 15    &        \\
\hline

\end{tabular}
\caption{\footnotesize Unigrams}
\label{tab:table1_a}
\end{subtable}
\label{table:ngram}
\end{table}

For a specific lecture session, the most frequent $N$-grams are examined one by one, to determine whether the corresponding word(s) can be matched with any of the discovered ``word'' clusters. There are cases that a cluster may partially match a trigram or bigram. If the unmatched part is a function word, e.g. ``linear combination'' versus ``linear combination of'', it is regarded as a case of match. If the unmatched part is a content word, e.g., ``divisor'' versus ``common divisor'', it is regarded as mismatch.


For Lecture 4 (``Number Theory I'') of the course MATH and Lecture 8 (``Object-Oriented Programming'') of the course PYTH, we analyze the top $10$ trigrams, top $20$ bigrams and $30$ unigrams and match them with the $10$ longest sequence clusters and other highly-populated clusters. The matching rates are found to be 
$73.3\%$ and $51.6\%$ respectively. The details of matching results for the lecture of ``Object-Oriented Programming'' are given as in Table \ref{table:ngram}. It is noted that the uncovered unigrams are mostly words with a small number of phones, e.g. ``add'', ``car'', ``code'', while the covered words are mostly polysyllabic words, e.g., ``python", ``coordinate''. For the lecture of ``Number Theory I'', the matching rate for unigrams is higher, due to more complicated phonetic structure of the words.

The same analysis has also been done for the course COMM, which contains $25$ lectures of $70$ minutes long. The $100$ most frequent trigrams, bigrams and unigrams are examined by comparing with clusters generated from all lectures. A high matching rate of $85.8\%$ is recorded.

The proposed system needs improvement in its ability of identifying keywords of relatively short length. In fact, short candidate sequences are not included when generating the candidate sequences for clustering (see Section \ref{ssec:bag_of_seq}). There is also an issue related to clusters that represent parts of a word.





\section{Conclusion}

We propose a two-stage approach to unsupervised keyword discovery. Pseudo transcription are generated by decoding results of unsupervised subword models.
Keywords are tokenized by searching matching subword sequences in pseudo transcription using local sequence alignment, and clustered into groups with leader clustering.

We experimented the model on 3 academic courses, concluding that the model has the ability in extracting topic related information, with coverage of $51.6\% - 85.8\%$ of the most frequent keywords. The performance of keywords matching is related to the phone size of the word, and is easier to match words with more phonemes with high purity in the clusters.

It is expected that the model has the capability in modeling other languages (e.g. zero-resource languages) with the use of language transfer property in multilingual bottleneck network. Future application on topic classification of recordings can also be considered.




\section*{Acknowledgment}
This research is partially supported by a GRF project grant (Ref: CUHK 14227216) from the Hong Kong Research Grants Council.

\bibliography{refs}
\bibliographystyle{IEEEtran}


\end{document}